\documentclass[10pt,letterpaper]{article}
\usepackage{graphicx}
\usepackage{amsmath}
\begin{document}
\title{Inverse design of nanophotonic structures using complementary convex optimization}
\author{Jesse Lu and Jelena Vu\v{c}kovi\'{c}\\\texttt{jesselu@stanford.edu}}
\date{E.~L.~Ginzton Laboratory, Stanford University, Stanford, CA 94305-4085, U.~S.~A.}
\maketitle

\begin{abstract}
A computationally-fast inverse design method for nanophotonic structures is presented. The method is based on two complementary convex optimization problems which modify the dielectric structure and resonant field respectively. The design of one- and two-dimensional nanophotonic resonators is demonstrated and is shown to require minimal computational resources.
\end{abstract}

\tableofcontents

% \ocis{350.4238 Nanophotonics and photonic crystals, 230.5750 Resonators, 230.5298 Photonic crystals, 220.0220 Optical design and fabrication}
\section{Introduction}
Numerous numerical methods have been devised to solve Maxwell's equations in both time\cite{Yee66} and frequency\cite{AB74,GA79} domains. We refer to these schemes as direct solvers, since they compute the electric and magnetic fields based on current sources, charge distributions and surrounding dielectric and/or metallic structures. While extremely useful in simulating optical components, using direct methods to design optical components, especially in two or three dimensions,  typically requires an extremely time-consuming brute force search in a large parameter space\cite{Lon09,Sch02,Nod05,Lip08,SD05}.

On the other hand, an inverse solver would be much more adept in such design and optimization problems\cite{Sig04,Vuc05}. In this work, we define the inverse problem as that where the electromagnetic field is known, but the surrounding structure is not known. The goal in the inverse problem is, then, to find a dielectric structure that will produce that specific electromagnetic field profile.

We show that one can design nanophotonic resonators by specifying the electromagnetic field and its desirable characteristics (such as cavity quality (Q) factor and/or mode volume) and then using an inverse solver to find the corresponding dielectric structure. We show that the inverse method used is not only computationally-fast, but is also able to optimize for multiple device characteristics and produce multiple resonances, both of which are very difficult using direct methods. 

\section{Numerical Setup}
We start from the time-harmonic eigenvalue equation 
\begin{equation}
\nabla \times \epsilon^{-1} \nabla \times H = \left(\frac{\omega}{c}\right)^2 H
\label{trad ev}\end{equation}
where $H$, $\epsilon$, $\omega$ and $c$ are the magnetic field, relative permittivity, resonance frequency and speed of light respectively. To solve the problem numerically, $H$ and $\epsilon$ are discretized in space using the standard Yee cell used in finite difference methods\cite{Yee66}. Also, the curl operators, since they are linear, are represented by the matrix $A$. Eq.~\eqref{trad ev} can now be written as 
\begin{equation}
A Y A x = \xi x
\label{lin ev}\end{equation}
where
\begin{align*}
&A \text{ is the discretized curl operator,} \\
&Y = \text{diag}(\epsilon^{-1}) \text{ is the diagonal matrix representing the dielectric structure,} \\
&x \text{ is a vector representing $H$, and} \\
&\xi =  \left(\frac{\omega}{c}\right)^2. 
\end{align*}
In this form, given $Y$, we can solve the direct problem by computing $x$ using an eigenvalue solver\cite{JJ99}. However, we note that eq.~\eqref{lin ev} is also linear in $Y$, which allows us, if $x$ is held constant, to solve the inverse problem by expressing eq.~\eqref{trad ev} as
\begin{equation}
By = d
\label{inv ls}\end{equation}
where
\begin{align*}
&B = A \cdot \text{diag}(Ax), \\
&d = \xi x\text{, and} \\
&y = 
\begin{bmatrix}
\epsilon_1^{-1} \\ \epsilon_2^{-1} \\ \vdots
\end{bmatrix} \text{ the variable for which we solve.}
\end{align*}
Here, $\text{diag}(Ax)$ is the matrix with the values of $Ax$ along the main diagonal and zeros elsewhere.

\section{Least-Squares Method in 1D}
\subsection{Least-Squares}
\begin{figure}[htbp]\centering
\includegraphics[width=\textwidth]{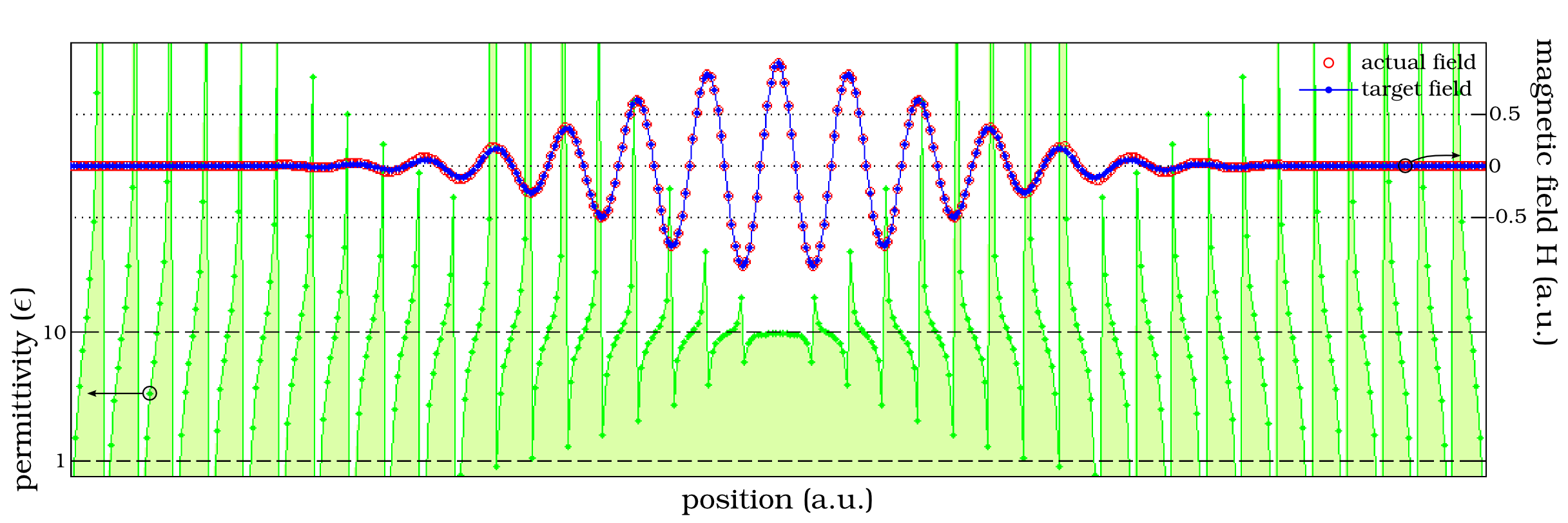}
\caption{Inverse design of a one-dimensional structure using the unmodified least-squares method. The target field is a sinusoid within a Gaussian envelope. The computed dielectric structure (green area) supports a field (red circles) that exactly matches the target field (blue line). The entire design process is also extremely fast and takes less than $1$ second to complete on a generic desktop computer. The periodic singularities in the dielectric structure are non-physical and will be addressed later in the article.}
\label{ls pic}\end{figure}
The result of applying this method to a simple one-dimensional problem is shown in Fig.~\ref{ls pic}. A generic least-squares solver~\cite{cholmod} was used to find the dielectric structure, $y$ (green region), that exactly produces the target field, $x$ (blue line), using eq.~\eqref{lin ev}. Using a generic desktop computer, the solution was obtained in less than a second. Then a finite-difference time-domain (FDTD) solver was used to obtain the actual field (red circles) produced by the structure and to verify the accuracy of $y$.  

As expected, Fig.~\ref{ls pic} shows that the target field is reproduced exactly by the dielectric structure. However, the resulting structure is full of undesireable singularities. The rest of the section focuses on producing a well-behaved dielectric structure that still reproduces the target field accurately.

\subsection{Regularized Least-Squares}
The simplest way to produce a well-behaved dielectric structure is to add a regularization term to our least-squares problem, 
\begin{equation}
\begin{bmatrix} B \\ \sqrt{\eta} I \end{bmatrix} y = \begin{bmatrix} d \\ \sqrt{\eta}y_0 \end{bmatrix}
\label{regls}\end{equation}
which is equivalent to solving the following optimization problem
\begin{equation}
\text{minimize}\quad \|By-d\|^2 + \eta\|y-y_0\|^2.
\label{regls alt} \end{equation}
Here $y_0$ represents some initial guess for the dielectric structure, that we want the values of $y$ to stay close to.

\begin{figure}[htbp]\centering
\includegraphics[width=\textwidth]{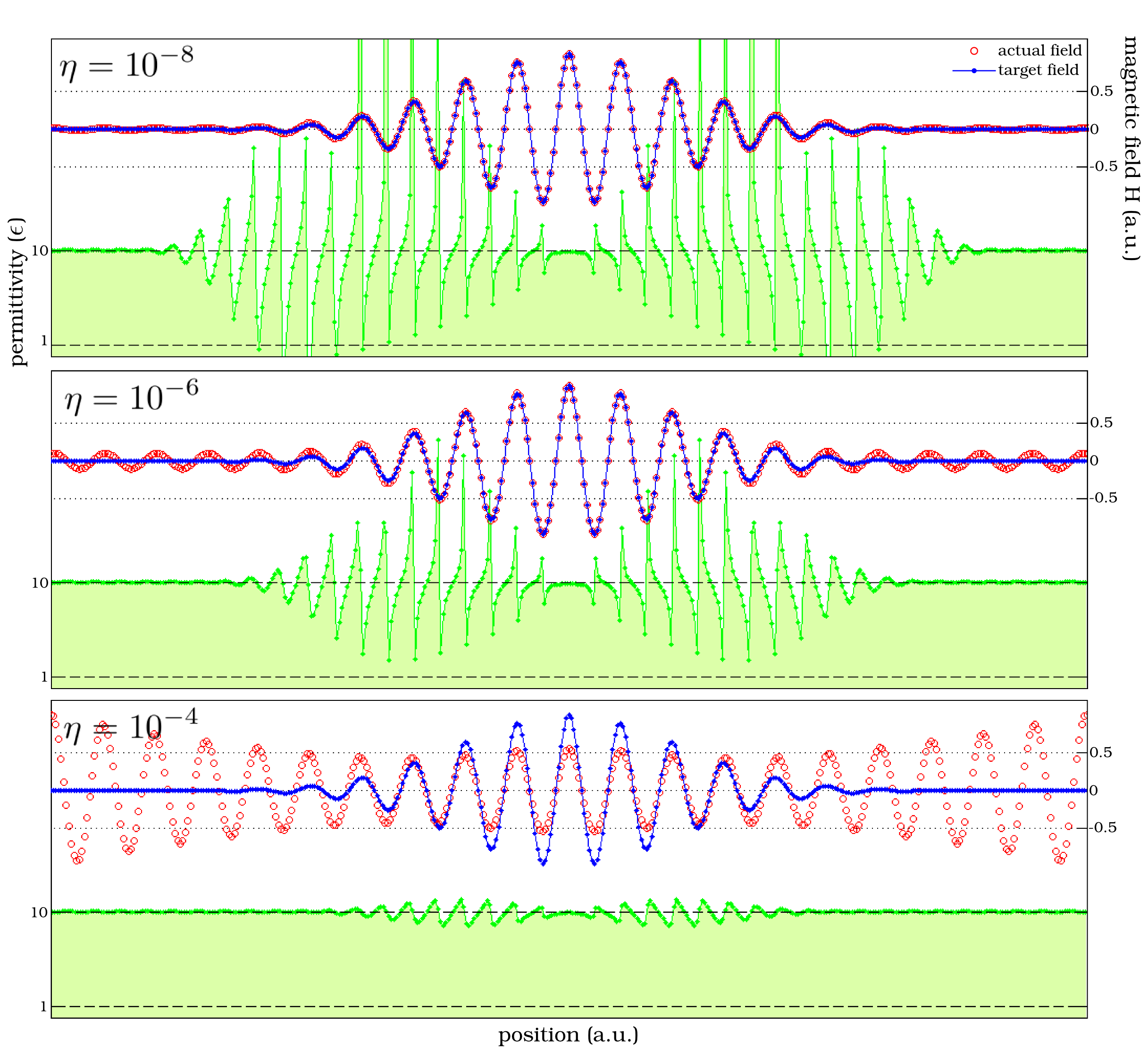}
\caption{Inverse design of one-dimensional structures using the regularized least-squares method. The same target field is used as in Fig.~\ref{ls pic}, and the computation time remains below $1$ second. As the regularization parameter, $\eta$, is increased, $\epsilon$ is increasingly constrained to a chosen constant value of $10$. At the same time, the mismatch between target and actual fields increases markedly. This illustrates the apparent trade-off between producing reasonable structures and accurately reproducing a fixed target field.}
\label{regls pic}
\end{figure}

This method allows us to trade off accuracy in order to keep $\epsilon$ close to acceptable values, by increasing $\eta$. We chose to constrain $\epsilon$ around a constant value of $\epsilon_0 = 10$ and solved the least-squares system for $\eta=10^{-8}$, $10^{-6}$, and $10^{-4}$. The results, each still obtained in under a second, are shown in Fig.~\ref{regls pic} and illustrate the trade-off between constraining $\epsilon$ and accurately reproducing $H$.

\section{Complementary Optimization in 1D}
\subsection{Motivation for a Complementary Optimization Strategy}
The fundamental problem in the previous examples is actually not in the methods themselves, but in the improper selection of a target field. In fact, it is very difficult to select a suitable target resonant field because not every resonant mode even has a corresponding dielectric structure that is able to reproduce it. Furthermore, it is nearly impossible to select a multi-dimensional field which corresponds to a well-behaved, isotropic and discretely-valued $\epsilon$, as would be needed for practical structures. 

For this reason, a successful method must be allowed to either modify the target field, or specify it completely, in which case the user would only determine certain characteristics (e.g. mode-volume, Q-factor) that the target field should have. The former strategy is developed in both one and two dimensions, while the latter strategy is implemented in Section \ref{sec:2Dbounded} in order to design two-dimensional resonators with discrete values of $\epsilon$.

\subsection{Complementary Optimization}\label{sec:comp}
We start with the same target field as in the previous examples but we now formulate a method that allows for it to be modified during the design process. The formulation chosen is a complementary optimization routine, where we continually alternate between modifying $\epsilon$ to better fit the field, and then modifying the field to better fit $\epsilon$. Here, we use the term ``fit'' to mean that either $\epsilon$ or $H$ is solved so that the residual error from eq.~\eqref{trad ev} is minimized. Additionally, both iterations are regularized in order to stably approach a solution. This algorithm can be summarized as follows, 
\begin{align}
&\text{choose $x_0$ and $y_0$} \nonumber \\
&\text{for } i = 1, 2, \ldots \nonumber \\
&\quad\text{minimize}\quad \|B_{i-1} y_i-d_{i-1}\|^2 + \eta_1\|y_i-y_{i-1}\|^2 \label{comp1}\\
&\quad\text{minimize}\quad \|A Y_i A x_i - \xi x_{i-1}\|^2 + \eta_2\|x_i-x_{i-1}\|^2 \label{comp2}
\end{align}
where $Y_i = \text{diag}(y_i)$, $B_i = A \cdot \text{diag}(A x_i)$ and $d_i = \xi x_i$. $\|A Y_i A x_i - \xi x_{i-1}\|^2$ is used instead of $\|A Y_i A x_i - \xi x_{i}\|^2$ to avoid the trivial $x_i = 0$ solution and does not affect the overall accuracy since $x$ changes very slowly. 

\begin{figure}[htbp]\centering
\includegraphics[width=\textwidth]{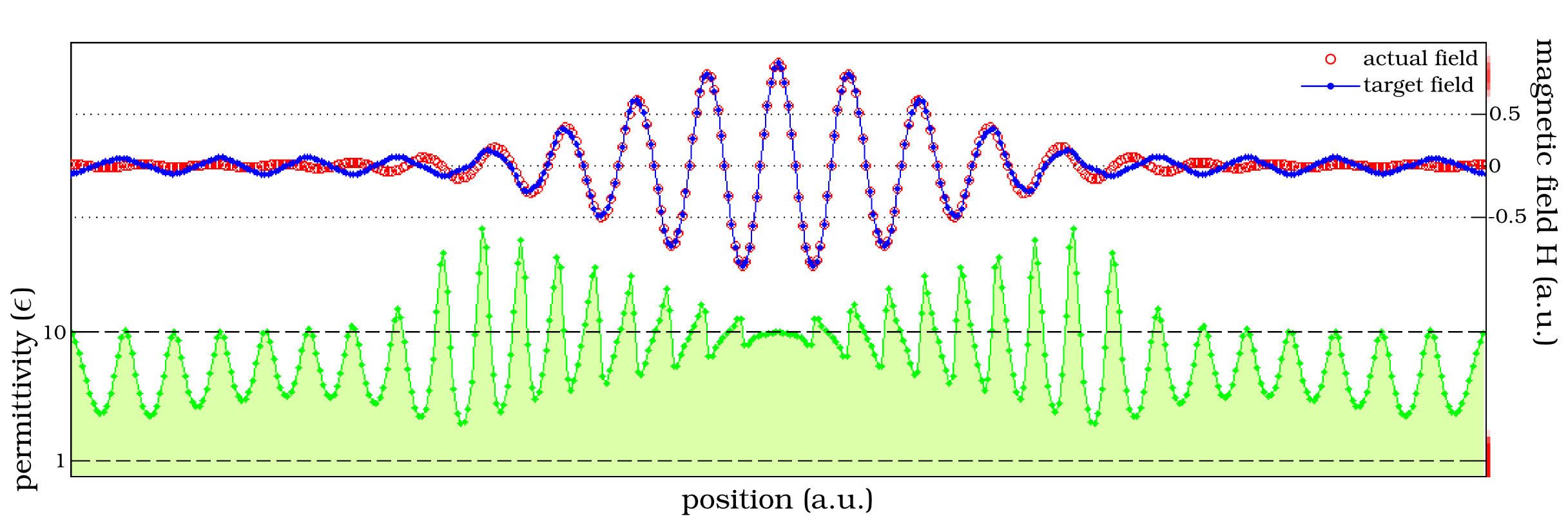}
\caption{Inverse design of a one-dimensional structure using the complementary optimization method. The target field in Figs \ref{ls pic} and \ref{regls pic} is used as the initial target field. The rates of change for both $\epsilon$ and $H$ are controlled by regularization parameters $\eta_1=10^{-4}$ and $\eta_2=10^{-3}$ respectively. The $400$ iterations used to achieve this result took $60$ seconds to compute. This method results in a well-behaved $\epsilon$ that actually produces a field very similar to the original target field. Interestingly, the formation of a ``steady-state'' periodic structure toward the sides of the structure has emerged.}
\label{comp pic}
\end{figure}

Fig.~\ref{comp pic} shows that the complementary optimization algorithm, after $400$ iterations and with the correct choice of regularization parameters $\eta_1$ and $\eta_2$, results in a well-behaved structure that is able to closely reproduce the modified target field. Numerically, the least-squares problem must now be solved numerous times, which increases the computational time needed to around $60$ seconds. 

\subsection{Complementary Optimization with Bounded $\epsilon$}
In order to achieve a more practical, discretely-valued dielectric structure, we can impose strict upper- and lower-bounds on $\epsilon$. To this end, we modify our algorithm as such,
\begin{align}
&\text{choose $x_0$ and $y_0$} \nonumber \\
&\text{for } i = 1, 2, \ldots \nonumber \\
&\quad\text{minimize}\quad \|B_{i-1} y_i-d_{i-1}\|^2 \nonumber \\
&\quad\text{subject to}\quad \epsilon_\text{max}^{-1} \leq y_i \leq \epsilon_\text{min}^{-1} \label{bc1} \\
\nonumber \\
&\quad\text{minimize}\quad \|A Y_i A x_i - \xi x_{i-1}\|^2 + \eta_2\|x_i-x_{i-1}\|^2. \label{bc2}
\end{align}
In this algorithm, eq.~\eqref{bc1} is a convex optimization problem\cite{BV04}. This allows us to impose hard constraints on $\epsilon$, which in turn allows us to remove the regularization term present in eq.~\eqref{comp1}. The \emph{CVX} package\cite{CVX} is used to solve eq.~\eqref{bc1}, with each iteration of the algorithm now requiring roughly $1$ second of computation time.
\begin{figure}[htbp]\centering
\includegraphics[width=\textwidth]{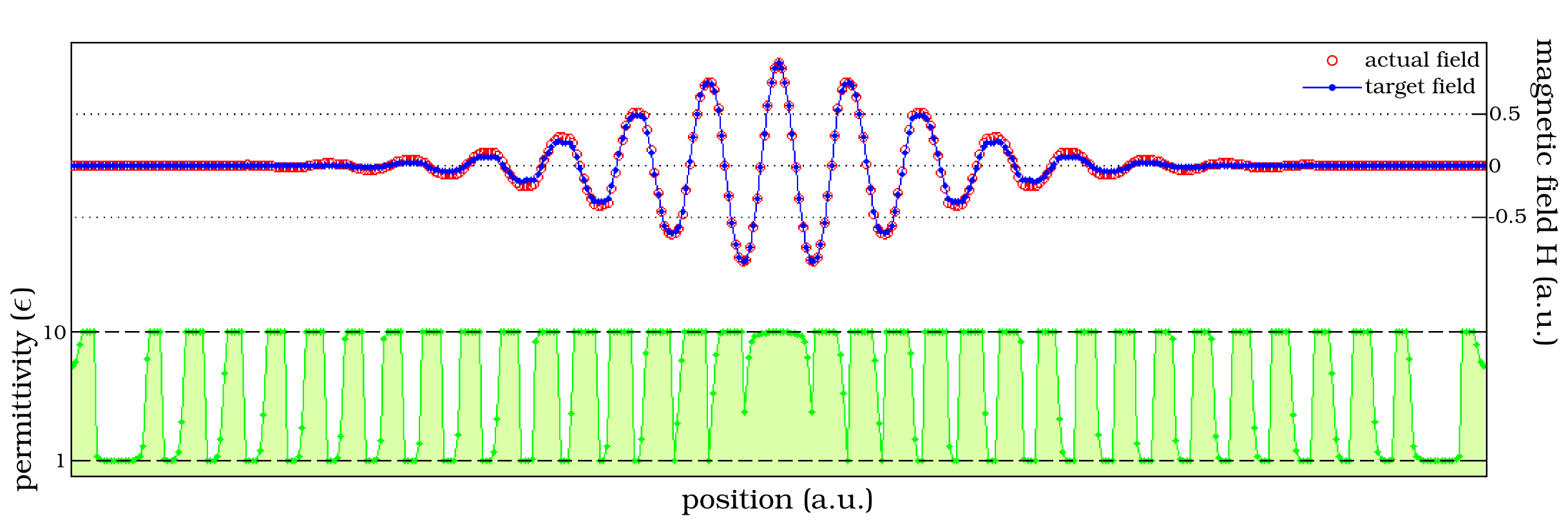}
\caption{Inverse design of a one-dimensional structure using the complementary optimization method with bounded $\epsilon$. The parameters are identical to those used to produce Fig.~\ref{comp pic} with the exception that only one regularization term is now needed ($\eta_2=10^{-3}$). The algorithm was run for $100$ iterations, which took $100$ seconds. The structure turns out to be almost completely binary-valued and looks like a periodic structure with tapered duty cycle. It produces an actual field which very closely matches the final target field.}
\label{bounded comp pic}
\end{figure}

A nearly binary-valued dielectric structure is obtained in Fig.~\ref{bounded comp pic}, which accurately produces the final target field. This is very useful for the design of practical structures, since they usually consist of two or three different materials at most. Interestingly, although the directly discreteness of $\epsilon$ was not enforced (since that would make the problem non-convex), a discrete, binary-valued structure has still arisen. 

\section{Complementary Optimization in 2D}
\subsection{``S'' Resonator}
We now demonstrate that the complementary optimization method is versatile and can be scaled to multiple dimensions. To ensure that $\epsilon$ is well-behaved we use a point-spread function which does not allow $\epsilon$ to change at a certain point in space without affecting the values surrounding it. 

\begin{figure}\centering
\includegraphics[width=\textwidth]{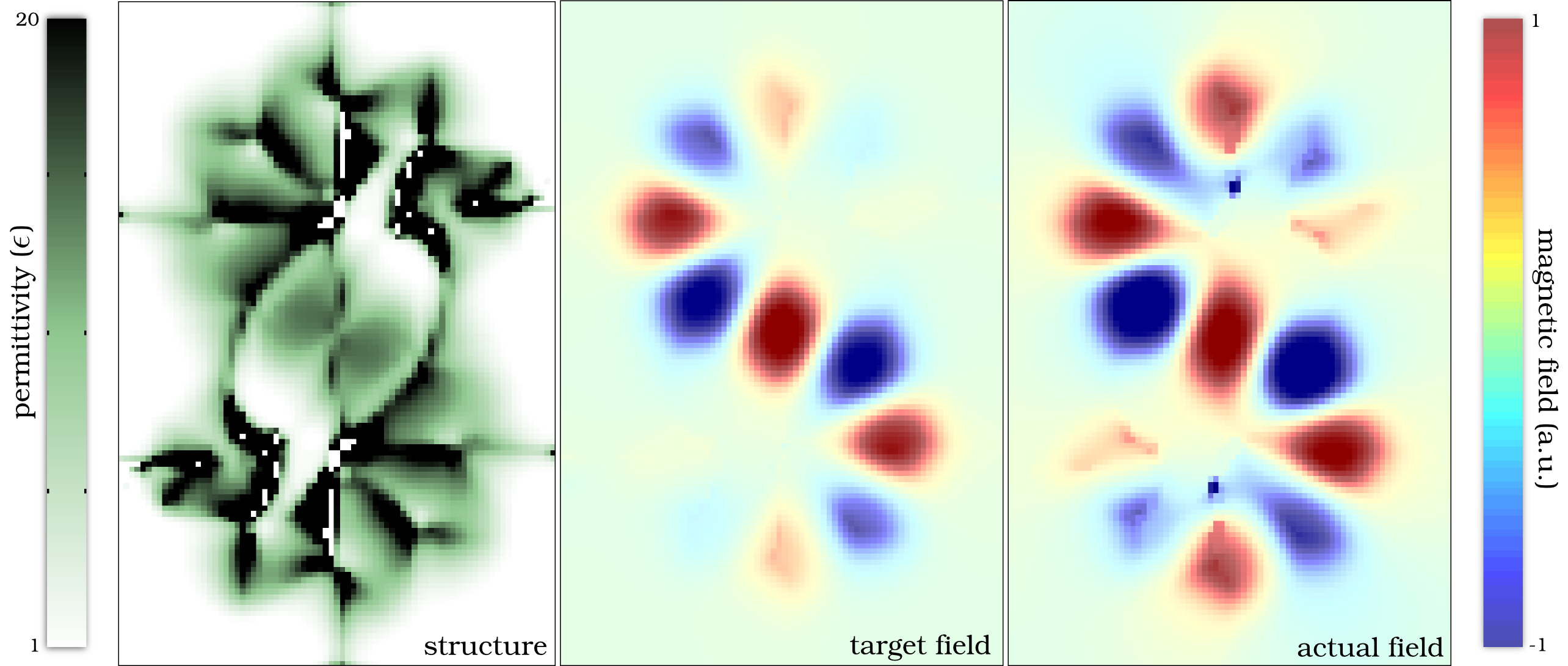}
\caption{Inverse design of an ``S'' resonator using the complementary optimization method without bounds on $\epsilon$. The design was initialized by specifying an initial dielectric structure ($\epsilon=1$ everywhere) and a resonant field in the shape of an ``S''. The final dielectric structure was produced after $50$ iterations which took $90$ seconds to complete in total. The final dielectric structure is quite unintuive, and yet reproduces the target field surprisingly well. This example demonstrates the versatility of the complementary optimization method in producing designs, from very simple specifications, which otherwise could be attained only with considerable difficulty.}
\label{S pic}
\end{figure}

In order to show that our method can produce complex designs, we choose an S-shaped target field which is non-trivial to reproduce. The optimization results, using the complementary optimization method from Section \ref{sec:comp}, are shown in Fig.~\ref{S pic}. The resulting dielectric structure is continuous, unbounded and contains some singularities (white dots), but the final target and actual fields match up well. Also, the computational cost remains quite reasonable; the $50$ iterations needed required only $5$ minutes of computation time. The resulting structure is completely unintuitive, and illustrates the kind of new capabilities offered by the inverse design strategy. Specifically, that a complex, intricate structure can be designed just by specifying the shape and frequency of a rather simple electromagnetic mode.

\subsection{Multi-Mode Inverse Design}
\begin{figure}\centering
\includegraphics[width=\textwidth]{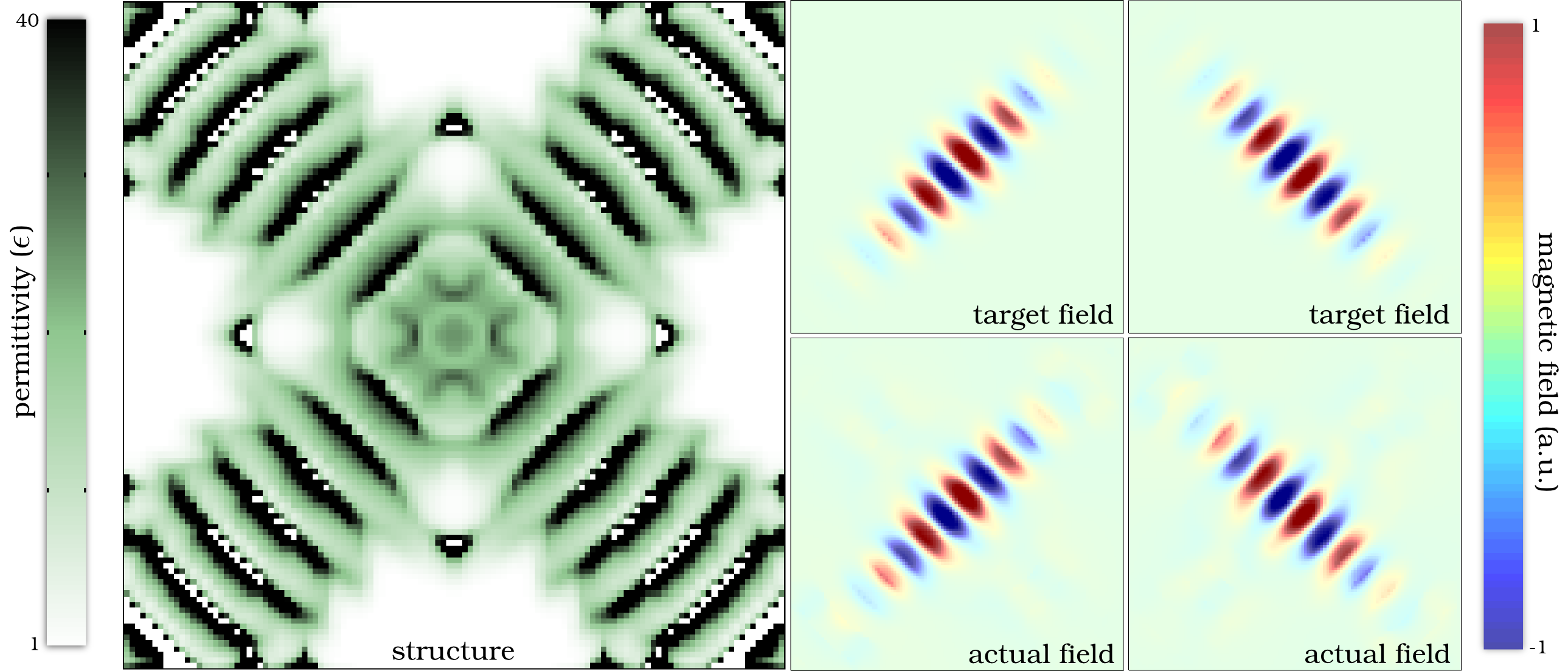}
\caption{Inverse design of a doubly-resonant, degenerate ``X'' resonator produced by the complementary optimization method with unbounded $\epsilon$. As in Fig.~\ref{S pic} the initial value of $\epsilon$ was $1$ everywhere. The two initial target fields used are only slightly perturbed and are very similar to the two final actual target fields. Computationally, this design took $5$ minutes to complete and required $40$ iterations. This example shows that the complementary optimization strategy can be extended to produce dielectric structures with multiple resonances. Such an ``X'' resonator is useful for polarization-entangled single-photon sources for example\cite{Hen06}.}
\label{X pic}
\end{figure}

The complementary optimization method can also be extended to produce dielectric structures with multiple resonances. To do so, multiple initial target fields are specified. The dielectric structure is first modified to simultaneously fit all target fields using a multi-objective least-squares method. Then each target field is individually modified to fit the structure; and we continue alternating between optimizing $\epsilon$ and $H^{(j)}$ in this way. A benefit of this scheme is that only the $\epsilon$ optimization increases in size, so the design process remains computationally tractable, even for several resonant fields. This algorithm can be summarized as follows,
\begin{align}
&\text{choose $y_0$, $x^{(1)}_0$, $x^{(2)}_0$, \ldots} \nonumber \\
&\text{for } i = 1, 2, \ldots \nonumber \\
&\quad\text{minimize}\quad \eta_1\|y_i-y_{i-1}\|^2 + \sum_j \|B^{(j)}_{i-1} y_i-d^{(j)}_{i-1}\|^2  \label{mm1}\\
&\quad\text{for } j = 1, 2, \ldots \nonumber \\
&\quad\quad\text{minimize}\quad \|A Y_i A x^{(j)}_i - \xi x^{(j)}_{i-1}\|^2 + \eta_2\|x^{(j)}_i-x^{(j)}_{i-1}\|^2. \label{mm2}
\end{align}
The design of an ``X'' resonator with two perpendicular, degenerate modes is shown in Fig.~\ref{X pic}. The added complexity increases the total computation time to $5$ minutes for $40$ iterations of the algorithm. 

\subsection{Design of Waveguiding Devices}
\begin{figure}[htbp]\centering
\includegraphics[width=\textwidth]{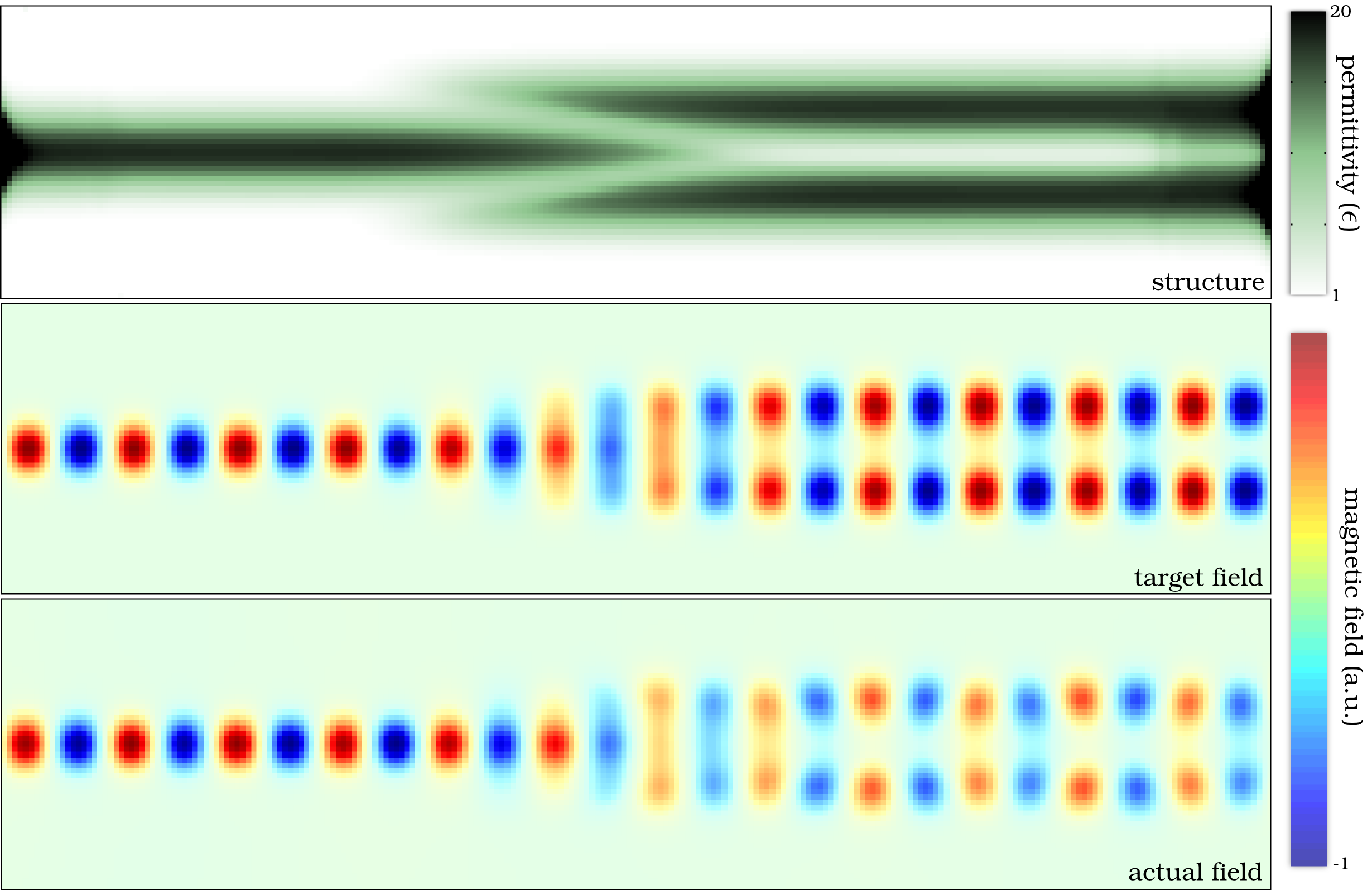}
\caption{Inverse design of a single-to-dual beam waveguide coupler using the complementary optimization method without bounds on $\epsilon$. Two degenerate modes with opposite symmetry (sine and cosine) are used as target fields. Only $4$ iterations ($14$ seconds) are needed to achieve this solution. This is a simple demonstration showing that the complementary optimization method can also be extended to design waveguiding devices.}
\label{wg pic}
\end{figure}

The multi-mode inverse design method can also be applied to the design of waveguiding devices such as multiplexers/demultiplexers, waveguide couplers, crossbars and dispersion-tailored waveguides. This can be accomplished by treating a waveguiding device as a doubly-degenerate resonator at its operational frequencies and then enforcing opposite symmetries (even/odd or cosine/sine) in the degenerate modes. A single-to-dual beam waveguide coupler designed based on Eq.~\eqref{mm1} and \eqref{mm2} is shown in Fig.~\ref{wg pic}, and motivates how one might design other waveguiding devices such as channel-drop filters. 
In addition to macroscopic waveguiding devices, one should also be able to create novel periodic waveguiding structures. For instance, by controlling the frequency of a particular waveguiding mode at each of its $k$-vectors, one could create a waveguide with a customized dispersion characteristic, which would be useful in the design of slow-light waveguides.

\section{Complementary Optimization with Bounded $\epsilon$ in 2D}\label{sec:2Dbounded}
\subsection{Numerical Method}\label{sec:2Dnum}
We now use the complementary optimization method to design resonators with discrete, binary $\epsilon$ in two dimensions. At the same time, the algorithm is modified so that we no longer specify an initial target field; instead, only an initial dielectric structure and the maximum desired mode volume (i.e. mode area in 2D) are specified. In addition, the optimization process attempts to maximize the Q-factor. Such an algorithm now consists of iteratively solving two convex optimization problem,
\begin{align}
&\text{choose $y_0$} \nonumber \\
&\text{for } i = 1, 2, \ldots \nonumber \\
&\quad\text{minimize}\quad \|A Y_{i-1} A x_i - \xi x_{i}\|^2 +   \eta \|F x_i\|^2\nonumber \\
&\quad\text{subject to}\quad (A x_i)^T Y_{i-1} (A x_i) \leq A_\text{mode} \label{be1} \\
\nonumber \\
&\quad\text{minimize}\quad \|B_i y_i-d_i\|^2 \nonumber \\
&\quad\text{subject to}\quad \epsilon_\text{max}^{-1} \leq y_i \leq \epsilon_\text{min}^{-1}. \label{be2} 
\end{align}
In this algorithm, the field optimization, Eq.~\eqref{be1}, differs significantly from Eq.~\eqref{bc2}. First, the eigenvalue equation term $\|A Y_{i-1} A x_i - \xi x_{i}\|^2$ no longer utilizes $x_{i-1}$. Also, a new ``Fourier-minimizing'' term $\|F x_i\|^2$ has been added. Here, the row vectors of $F$ consist of the field Fourier components that we do not want incorporated in the final solution. The motivation for adding this term is to design high-Q resonators using planar structures, where the Q-factor is limited by out-of-plane losses, and can thus be improved by eliminating field Fourier components that are not localized by total internal reflection (components inside the light cone)~\cite{Vuc05}. Lastly, the $(A x_i)^T Y_{i-1} (A x_i)$ term allows us to specify the mode area (mode volume in three dimensions), that we desire for our resonant field. This works because the two minimization objectives in Eq.~\eqref{be1} will generally cause the mode area to be as large as possible, which means we always end up with $(A_1 x_i)^T Y_{i-1} (A_1 x_i) = A_\text{mode}$.

The addition of the two new terms in Eq.~\eqref{be1} signifies that the field iteration in our algorithm attempts to do more than to just satisfy Maxwell's equations. Rather than only trying to fit the dielectric structure, the resonant field now also minimizes some of its Fourier components while working with a limited mode area. 

In contrast with the field optimization given by Eq.~\eqref{be1}, the structure optimization given by Eq.~\eqref{be2} remains identical to the equation in the one-dimensional case, Eq.~\eqref{bc1}, and contains no terms related to notions of Fourier components or mode area. In fact, the only objective in the $\epsilon$ optimization is to better fit the field generated from the field optimization. This means that in this scheme, the field iteration is leading the structure iteration. In other words, Eq.~\eqref{be1} ``looks ahead'' and gradually adapts itself to become a more desirable field, while Eq.~\eqref{be2} just ``follows along''. For this reason, this algorithm is unique from the other complementary optimization algorithms previously presented in the article, since in those algorithms both iterations follow one another and eventually ``meet in the middle''.

Finally, in this scheme, we chose to limit the degrees of freedom of both $x_i$ and $y_i$. Some components of $x_i$ were fixed at non-zero values and were not allowed to be modified simply to avoid the $x_i = 0$ situation. To enhance the aesthetics of the resulting dielectric structure, some components of $y_i$ were frozen as well, which is useful in cases where one would only like to modify the dielectric structure within a waveguide only, for example. 

\subsection{Circular Grating Resonator}
\begin{figure}[htbp]\centering
\includegraphics[width=\textwidth]{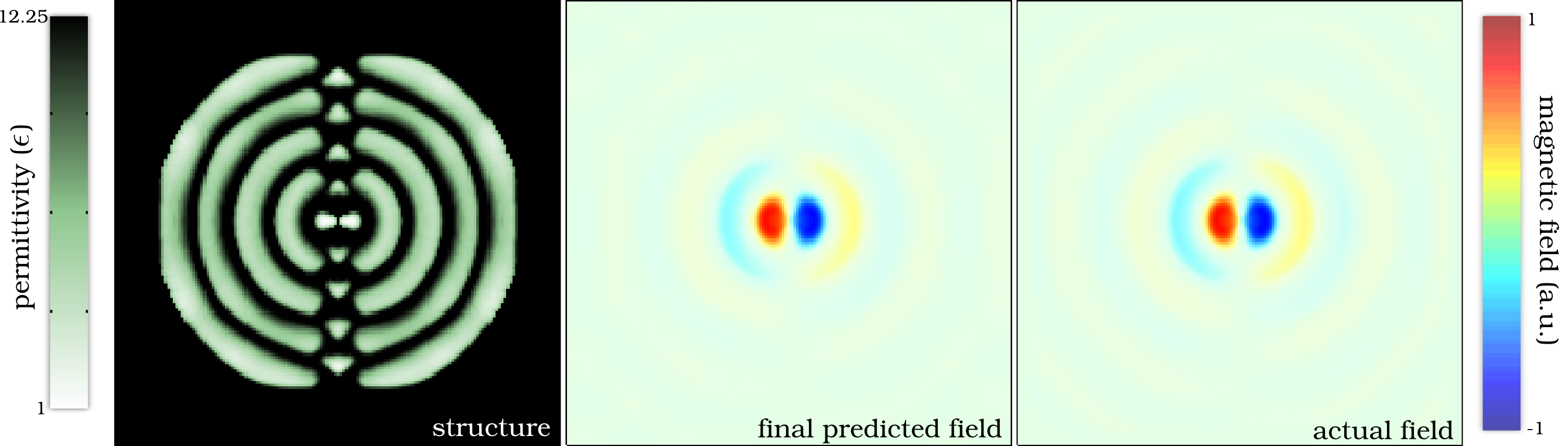}
\caption{Inverse design of a two-dimensional resonator using the complementary optimization method with strict bounds on $\epsilon$. The initial specification is very simple and consists an initial dielectric structure ($\epsilon=12.25$ everywhere), the frequency and mode-volume of the resonance field as well as a weighting factor, $\eta$, to avoid leaky field Fourier components. Additionally, the values of $\epsilon$ are only allowed to be modified within a central circular region and must be kept between $1$ and $12.25$. After $40$ iterations which took $7$ minutes to complete, a discrete structure emerged with excellent match between the predicted ($x_{40}$) and actual fields. The structure resembles a circular grating with a bowtie-like central structure for focusing the resonant energy to a single point.} 
\label{circle pic}
\end{figure}
Fig.~\ref{circle pic} shows the design of a circular grating resonator using the method from Section \ref{sec:2Dnum}. The dielectric structure emerged from a very simple choice of initial structure, namely a constant $\epsilon=12.25$ everywhere. The range of $\epsilon$ was limited to be from $1$ to $12.25$, the mode area was set to $5$ and $\eta= 10^{-3}$. Central components of $x_i$ were held constant to ensure that an electric dipole in the y-direction was produced in the center of the structure. Additionally, the components of $\epsilon$ outside a specified circle were held at a constant $\epsilon=12.25$ for the duration of the design process. The entire algorithm was run for $40$ iterations and took $7$ minutes.

After $40$ iterations, we see that a strong binary-valued structure has formed. Interestingly, the central bowtie-like structure has emerged from previous geneetic optimization methods as well\cite{Lip08}. Note also the extremely small amount of information needed to produce this result: only the frequency and mode area of the resonant field desired, and a trivial intial $\epsilon$. The ability to produce a rather advanced design from such a simple problem specification highlights the potential usefulness of inverse methods in the design of novel nanophotonic devices.

\subsection{Beam Resonator}
\begin{figure}[htbp]\centering
\includegraphics[width=\textwidth]{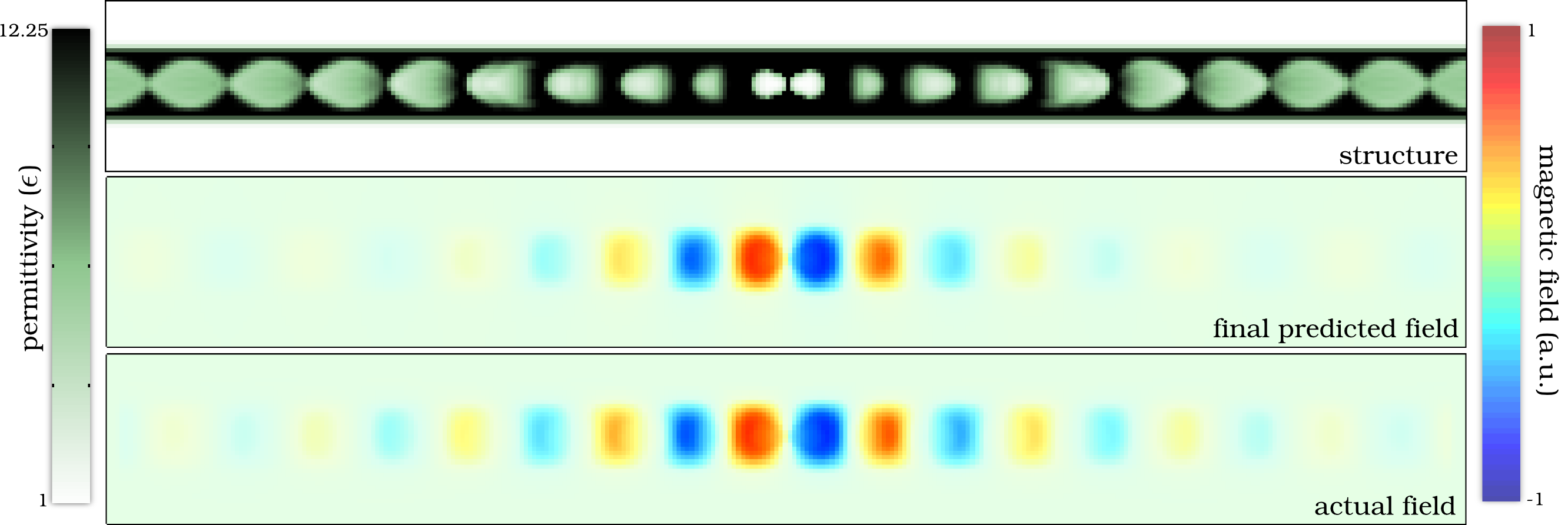}
\caption{Inverse design of a beam resonator in two dimensions using the complementary optimization method with bounded $\epsilon$. The initial conditions are identical to those for Fig.~\ref{circle pic}, except that the initial dielectric structure is an unbroken waveguide, and $\epsilon$ can only be modified within that waveguide. The structure emerged after $40$ iterations, which took $5$ minutes of computation. The bowtie-like structure has reappeared in the center. Interestingly, the effect of the Fourier term in Eq.~\eqref{be1}, is seen in the outward tapering of the holes.}
\label{line pic}
\end{figure}

The same approach was used to design a beam resonator as shown in Fig.~\ref{line pic}. The parameters used in this design are identical to those used for the circular resonator, with the exception that the initial dielectric structure now consists of an unbroken waveguide, and the region where $\epsilon$ can be modified is now confined to the center of the waveguide.

These two, simple modifications produce a vastly different result as is plainly seen from Fig.~\ref{line pic}. Some characteristics remain, such as the two closely spaced holes in the center of the structures which focus the electromagnetic energy to a central point. Other features are unique to the beam resonator, such as a gradual outward tapering of hole diameters and size ending in a periodic ``steady-state'' structure, in similar fashion to Fig.~\ref{comp pic}. This tapering is a direct result of the Fourier term in Eq.~\eqref{be1} as it leads to a smooth field profile variation and minimization of the components in the light cone that lead to out-of-plane losses\cite{Vuc05,Aka05}.

\section{Conclusions and Outlook}
In summary, we have introduced and demonstrated a complementary optimization method for the inverse design of nanophotonic resonators as well as waveguiding structures. We have demonstrated that, in two dimensions, this method can be used to design non-trivial as well as multi-resonant structures. Furthermore, we have shown that by constraining the range of available dielectric constants, the same underlying method can produce binary-valued, discrete dielectric structures. 

The methods presented here can readily be extended into three dimensions. In essence, the only hurdle is the size of the convex optimization problem for the field optimization. Since a full three-dimensional grid often consists of tens of millions of variables, solving the field optimization problem, Eq.~\eqref{be1}, using methods such as LU factorization is no longer computationally tractable. Instead, iterative methods such as truncated Newton methods must be employed. As it turns out, such iterative methods are especially well suited to this complementary optimization scheme, since they can take advantage of the information in $x_{i-1}$ in order to solve $x_i$ very quickly. This is very applicable to our algorithm since the field changes very little from one iteration to the next. 

Although solving the field optimization problem is numerically challenging in three dimensions, solving the structure optimization problem, Eq.~\eqref{be2}, in three dimensions is much simpler. This is because only structural variations in-plane need to be accounted for planar nanophotonic devices. The number of variables in Eq.~\eqref{be2} for both two- and three-dimensional problems is then effectively the same, and the same numerical techniques can be employed to solve both.

A full three-dimensional inverse design method based on complementary optimization would enable the design of novel nanophotonic resonators. For instance, a resonator for efficient wavelength-conversion could be designed by creating a structure with multiple well-placed resonant frequencies~\cite{Riv09}. Furthermore, effects such as mode overlap and the use of frequency-dependent refractive index materials could be taken into account with such a method.

\subsection*{Acknowledgements}
This work has been supported in part by the AFOSR for Compex and Robust On-chip Nanophotonics (Dr. Gernot Pomrenke).
We thank Prof. Stephen Boyd for many fruitful discussions regarding convex optimization.

\end{document}